\documentclass[sigconf]{acmart}
\acmConference[ESEC/FSE 2023]{The 31st ACM Joint European Software Engineering Conference and Symposium on the Foundations of Software Engineering}{11 - 17 November, 2023}{San Francisco, USA}
\AtBeginDocument{%
  \providecommand\BibTeX{{%
    \normalfont B\kern-0.5em{\scshape i\kern-0.25em b}\kern-0.8em\TeX}}}

\begin{document}

\title{Fault Injection based Failure Analysis of three CentOS-like Operating Systems}

\author{Hao Xu}
\email{2131483@tongji.edu.cn}
\affiliation{%
  \institution{Tongji University}
  \country{China}}

\author{Yuxi Hu}
\email{huyuxi.pt@alibaba-inc.com}
\affiliation{%
  \institution{Alibaba Inc.}
  \country{China}}

\author{Bolong Tan}
\email{bolong.tbl@alibaba-inc.com}
\affiliation{%
  \institution{Alibaba Inc.}
  \country{China}}

\author{Xiaohai Shi}
\email{xiaohai.sxh@alibaba-inc.com}
\affiliation{%
  \institution{Alibaba Inc.}
  \country{China}}

\author{Zhangjun Lu}
\email{2131482@tongji.edu.cn}
\affiliation{%
  \institution{Tongji University}
  \country{China}}

\author{Wei Zhang}
\email{1910134@tongji.edu.cn}
\affiliation{%
  \institution{Tongji University}
  \country{China}}

\author{Jianhui Jiang}
\authornotemark[1]
\email{jhjiang@tongji.edu.cn}
\affiliation{%
  \institution{Tongji University}
  \country{China}}


\begin{abstract}
The reliability of operating systems (OSs) has always been a focus of attention in both academia and industry. This paper presents a novel methodology for failure analysis of Linux-like OSs based on fault injection. Initially, we systematically define Linux-like fault modes by adopting the method of fault mode generation based on functional module division of Linux-like OSs. Subsequently, we construct a Linux fault mode library and develop a fault injection tool based on the fault mode library (FIFML). Finally, we conduct fault injection experiments on three commercial Linux distributions, i.e. CentOS, Anolis OS and openEuler. To reasonably divide the influence level and reduce the impact of performance fluctuations, we introduce three performance metrics including performance threshold, performance standard deviation, and the worst performance. Additionally, we employ failure rate, performance degradation rate, and performance level after fault injection to quantitatively describe the influence of fault injection on OS performance. Utilizing these metrics, we measure the performance disparity of three OSs. The experimental results show that Anolis OS outperforms CentOS and openEuler in virtual file systems, network interfaces, and process management systems. These findings underscore the significance of our methodology in assessing OS reliability. By comprehensively examining various fault modes and their effects on performance, our methodology contributes to a better understanding of OS failure behavior and provides insights for future system optimization.
\end{abstract}

\ccsdesc[500]{General and reference~Reliability}
\ccsdesc[500]{Software and its engineering~Software testing and debugging}
\keywords{CentOS, Anolis OS, openEuler, failure analysis, fault injection}


\maketitle

\section{Introduction}
A failure in any component of a software system can result in system failure, which negatively impacts the user experience. The operating system (OS) should ideally maintain good quality of experience even in the presence of faults, such as isolating a failed component without compromising system reliability and responsiveness. Reliability is an essential attribute of product or system quality, and failure analysis plays a crucial role in identifying the causes of system reliability problems. It helps identify failure modes, failure causes, and failure mechanisms, and propose measures to prevent future failures, thus improving the system's reliability.

Fault injection techniques are crucial for identifying and analyzing system failures in a controlled environment. By intentionally causing failures, developers can study the chain of events leading up to system failures, identify root causes, and improve the system to prevent future failures. However, developers are often hesitant to implement fault injection techniques due to the growing complexity and cost of modern systems, particularly when it comes to deciding which faults or errors to inject. In this context, failure analysis based on fault injection has emerged as an effective means to analyze system reliability, providing a valuable tool for identifying and addressing system failures in software systems.

Various researchers have studied system reliability by injecting faults into different components of the system. For example, Amarnath \cite{b1} and Winter \cite{b2} have injected faults into CPU registers and drivers, while Yoshimura et al. \cite{b3} have injected faults into application processes to investigate the propagation of errors to the kernel. Arlat \cite{b4} et al. have analyzed the representativeness of OS fault injection and compared the similarities and differences between real faults and injected faults using Linux as an example. Since the 1990s, academia has published numerous achievements in the development and application of fault injection tools, including fault injector in FTAPE \cite{b18}, fast fault injection FSFI \cite{b21}, and OS fault injection tools such as SockPFI \cite{b19} and TFI \cite{b20}. However, the majority of these works suffer from issues related to the application or coverage of their analysis methods, such as coarse granularity, inaccuracy, and low coverage. Additionally, some fault injection tools for OSs have been developed by industry, including FailViz \cite{b26} and DICE \cite{b27}, but these tools have limitations such as poor expandability, inaccurate injection, and poor authenticity.

This paper aims to present a methodology for failure analysis of Linux-like OSs based on fault injection. Specifically, we provide OS developers with an effective method for defining, executing, and analyzing faults, as well as a fault injection tool. In this paper, we focus on CentOS 8, Alibaba's Anolis OS, and Huawei's openEuler, and provide examples at each step to help readers better understand the process or adopt these steps in practical scenarios. Our contributions include: 1) We performed fault injection on internal functions at the kernel level and constructed a fault mode library for Linux. We also provided a systematic, lightweight guidelines for defining fault models through OS’s function modules, which is applicable to large software systems. 2) We developed a fault injection tool based on the fault mode library (FIFML) that enables the implementation of fault models through reusable fault injection plugins and an extensible architecture. 3) We conducted comparative experiments on CentOS 8, Anolis OS and openEuler distributions to analyze their failures and identify deficiencies in terms of reliability. We also compared the performance of these three open-source OSs.

The remainder of this paper is organized as follows. In Section 2, we introduce the methodology for failure analysis of Linux-like OSs based on fault injection, which includes key concepts and our proposed fault injection method based on internal function. Section 3 presents the failure analysis method based on fault injection, the fault mode generation method based on functional module division and the Linux fault mode library we have built. In Section 4, we detail the design of the fault injection tool FIFML and its verification for validity, completeness, and authenticity. The Section 5 describes the design and analysis of fault injection experiments on three OSs, as well as a performance analysis experiment. Finally, in Section 6, we conclude this paper.

\section{Methodology for Failure Analysis of Linux-like OSs based on Fault Injection}

\subsection{Background}
The goal of our failure analysis is to evaluate the impact of failures on Linux-like OSs. Evaluating system credibility and verifying the correctness of fault-tolerant mechanism design and implementation are crucial steps in the developing dependable computer systems. Artificially manufacturing faults in a verified system, by replacing natural physical faults with artificial faults, can effectively shorten the verification cycle. Fault injection is a process that involves using artificial methods to generate faults based on a selected fault type. The generated faults are then applied to the target system to accelerate its error process. The fault response information is collected and analyzed in a timely manner, and the final results are provided to technical personnel for research purposes. Therefore, our approach is to inject faults in the internal functions of the kernel and analyze the responsiveness and availability of the whole system in the presence of the injected faults. A key part of this approach is the generation of failure modes, since failures can generate from any hardware and software component of the OS. Since the Linux-like OSs is a large, complex software system, this method emphasizes software faults injection, which is the main cause of reliability problems in such systems.

In software operation, a state that fails to provide the expected function is called a software failure, and failure analysis is an essential method for improving software reliability. The cause that has the potential to cause failure is defined as a fault, which can also be described as the adjudged or hypothesized cause of an incorrect state of the system \cite{b5}. Software faults can originate from open source OS development or from proprietary customizations introduced by developers. The approach performs a series of fault injection experiments, where each experiment simulates a fault in an internal function of the kernel.

The Linux fault mode library is constructed by analyzing (including theoretical and experimental analysis), abstracting and summarizing the performance of possible failures in the Linux systems. Fault injection for Linux-like OSs is based on the Linux fault mode library, we select a single fault mode or combines multiple fault modes, generate a fault injection scheme, and forms a fault instance according to the conditions of the target system and its operating environment for implementation. Fault injection in Linux can accurately simulate kernel faults at the function level. Linux fault injection implemented by simulation include kernel function fault simulation, delay simulation, buffer data error simulation, system downtime restart simulation, kernel denial of service simulation, CPU usage simulation, etc.
\begin{figure}
\centering
\includegraphics[width=0.40\textwidth]{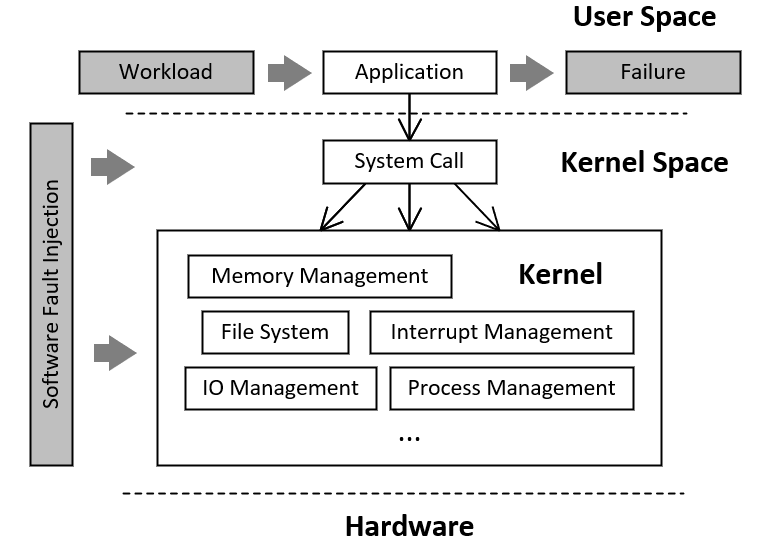}
\caption{\label{fig:fig1}General Structure of Linux}
\end{figure}
\subsection{Fault Injection based on Internal Functions}
The current main measures of fault injection are orthogonal defect classification (ODC) based fault injection \cite{b6} and interface-based fault injection \cite{b28}. The faults injected by ODC based fault injection are usually dormant, which leads to inefficiency because most of them are difficult to trigger. In contrast, interface-based fault injection does not have these problems. Its principle is to simulate the impact of faults in components by injecting exceptions or invalid values into their interfaces. Interface-based fault injection is a practical and popular approach that does not require any changes to the component's code, and since each fault injection generates errors, the efficiency of experimentation is ensured. However, our fault injection method based on internal functions has all the advantages of interface-based fault injection. Moreover, due to the further analysis of the internal functions, this fault injection method is more accurate and complete, and has wider coverage.

The principle of fault injection based on internal function is to inject faults into the layer below the target level, and the impact is observed at the target level to extract the fault mode. The Linux system consists of four parts, namely hardware, Linux kernel, Linux services, and user applications, as shown in Figure1. Among them, the Linux kernel is mainly used for abstraction and access scheduling of hardware resources. Fault injection for the Linux should be performed at the system call layer if it is based on the interface and at the kernel function layer if it is based on internal functions. When a Linux kernel code defect is activated, it will cause a Linux runtime fault, and the fault will propagate to the system call layer, causing an error when interacting with the upper-layer applications of Linux, and may even cause some programs to fail. The Linux kernel defect-fault-failure propagation mechanism is shown in Figure2.
\begin{figure}
\centering
\includegraphics[width=0.4\textwidth]{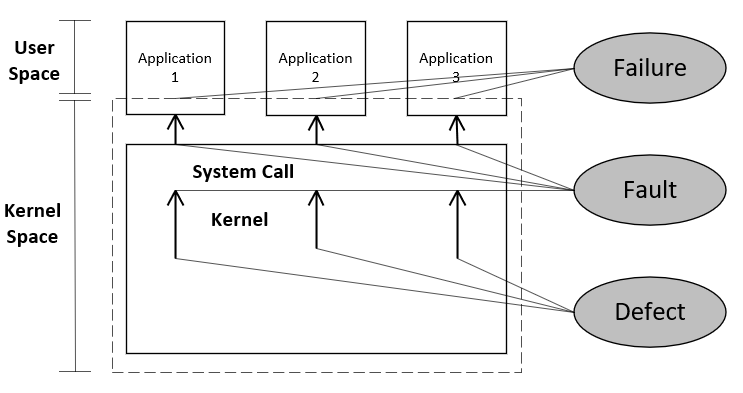}
\caption{\label{fig:fig2}Linux Kernel Defect-Fault-Failure Propagation Mechanism}
\end{figure}
To provide a more specific example, the system call corresponding to “$Set$ $memory$ $permission$” is mprotect(). The function description of mprotect() is as follows:
\[
SYSCALL\_DEFINE3(mprotect,\ unsigned\ long, start,
\]
\[
\ size\_t,\ len,\ unsigned\ long,\ prot)
\]

mprotect() calls the function do\_mprotect\_pkey(), which is the target of fault injection. The function description of do\_mprotect\_p\\key() (The pkey value is fixed to -1 when mprotect() is executed) is as follows:
\[
static\ int\ do\_mprotect\_pkey(unsigned\ long\ start,
\]
\[
size\_t\ len, unsigned\ long\ prot, int\ pkey)
\]
The analysis of do\_mprotect\_pkey() is as follows:
\begin{enumerate}
\item If the first parameter start is wrong, the fault mode of “invalid address parameter error when setting memory permission” may occur. When memory access permission is set, start is an invalid pointer, or is not an integer multiple of the page size. It is not aligned with the memory page. The function returns -EINVAL for “Invalid argument”.
\item If the second parameter len is wrong, the fault mode of “memory overflow error when setting memory permissions” may occur. When memory access permission is set, the specified address space [start, start+len-1] exceeds the actual process address space ( this is also related to the start parameter). The function returns -ENOMEM for “Out of memory”.
\item If the third parameter prot is wrong, the fault mode of “permission conflict error when setting memory permissions” may occur. After the mmap() is used to map a read-only file to memory, if you try to give this memory PROT\_WRITE permission (write permission), the service will be denied due to a permission conflict error. The function returns -EACCES for “Permission denied”.
\end{enumerate}

\section{Failure Analysis based on Fault Injection and Fault Mode Generation based on Functional Module Division}
\subsection{Failure Analysis Process and Methods}
When analyzing the failure of a Linux system, Linux can be divided into various abstraction levels, such as kernel functions, system calls, and user applications. The Linux kernel mainly abstracts and schedules hardware resources. The high complexity of Linux makes it challenging to generate fault modes as different functional modules behave differently after a fault occurs. Faults, errors, or failures in the runtime of Linux are typically caused by code defects being activated, hardware errors, even Linux application errors. Performing failure analysis directly on Linux code is time-consuming and laborious.

To address these challenges, we established a failure analysis process for Linux-like OSs based on fault injection as shown in Figure 3. We have previously used ODC and software-implemented fault injection \cite{b6} to perform failure analysis on the Linux kernel code. The experimental results indicate that 59.66$\%$ of kernel code are related to the Linux system call chain, which covers nearly 60$\%$ of Linux kernel code failures. Therefore, we adopt the fault mode generation based on functional module division to generate Linux fault modes and construct a Linux fault mode library to support Linux failure analysis based on fault injection.
\begin{figure}
\centering
\includegraphics[width=0.25\textwidth]{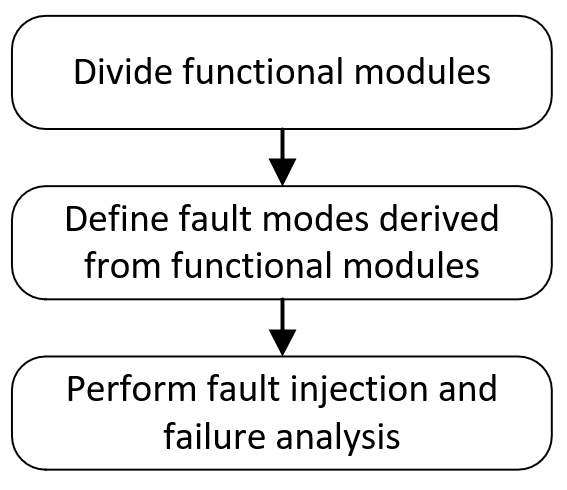}
\caption{\label{fig:fig3}Process of Failure Analysis based on Fault Injection}
\end{figure}

\textbf{Step 1: Divide functional modules.}

In this step, we first divide the target software (Linux-like OS) into functional modules, i.e. file system, interrupt management, IO management, memory management, and process management. After obtaining the large functional modules, we identify the key functions of each module (such as setting memory permissions, sending signals to processes, opening files, managing semaphore sets, sending messages, etc.). This process needs experience to ensure the quality of the subsequent fault mode extraction. It is not necessary to fully understand the internal details of the module, but it is necessary to master the interface and parameter information.

\textbf{Step 2: Define fault modes derived from functional modules.}

Failure scenarios are used in Linux failure analysis to help find the direct causes of these failures. The types of failure scenarios we consider are defined according to the relevant literature \cite{b14-17} and our previous work on modeling the effects of software failures \cite{b8-11}, including:

\begin{enumerate}
\item Kernel function failure: The return value of a system call is wrong, and it is unable to provide correct service.
\item Delay: The long delay makes it unable to provide services within the user's waiting time.
\item Buffer data error: The buffer data is wrong, and it cannot provide the correct service.
\item System downtime and restart: The system breaks down or restarts, and it cannot provide services.
\item Kernel Denial of Service: The system kernel refuses to provide services to users.
\item System CPU usage increases: The increase in CPU occupancy affects the services provided to users.
\end{enumerate}

By combining past experiences with fault injection and analyzing functional issues in Linux, we can identify fault modes in a Linux-like OS. Although this process may not uncover the root cause of the failure, it is a comprehensive, convenient, and effective way to identify issues. The list of issues includes the following:
\begin{enumerate}
\item Does the function return an abnormal result due to invalid parameters or values outside the specified range of the called kernel function? If yes, then one or more failure modes corresponding to “kernel function failure” are generated.
\item Does the function return an abnormal result due to insufficient privileges or the wrong address of the called kernel function? If yes, then one or more failure modes corresponding to “kernel function failure” are generated.
\item Is the function delayed or unresponsive during execution? If yes, then one or more failure modes corresponding to “delay failure” are generated.
\item Does the function produce an error result of unknown type or return incorrect results due to external data errors? If yes, then one or more failure modes corresponding to “buffer data error” are generated.
\item Does the function cause data synchronization errors or abnormal data due to unexpected system shutdown and restart? If yes, then one or more failure modes corresponding to “system shutdown and restart” are generated.
\item Does the function generate an exception due to the process being killed? If yes, then one or more failure modes corresponding to “kernel denial of service” are generated.
\item Does the function affect the service due to the system overload? If yes, then one or more failure modes corresponding to “system CPU usage failure” are generated. 
\end{enumerate}

We obtain the corresponding fault modes by analyzing related functions. It is necessary to consider the system calls and related parameters used by the functions. The fault modes are generated based on characteristics of system call and kernal functions, and added to the fault mode library. This process is repeated until no any new fault mode is generated, then the construction of the fault mode library is finished.

\textbf{Step 3: Perform fault injection and failure analysis.}

The fault injection tool injects faults into the OS. After a fault is injected, it is necessary to track the propagation process of errors generated by the injected fault, monitor the behavior of the OS, and analyze the failure caused by the injected fault. The fault injection tool must accurately inject or simulate the fault modes, and the OS source code must not be changed during the injection process.

\subsection{Examples of Linux Fault Mode Generation}
The analysis and collection process of the Linux fault modes served by the “$Manage$ $Semaphore$ $Set$” operation is given below. According to the list, the function may occur “kernel function failure”. The system call that corresponding to this operation when Linux is running is semop(). The function description of semop() is as follows:
\[
SYSCALL\_DEFINE3(semop,\ int,\ semid,
\]
\[
struct\ sembuf\ \_\_user\ *,\ tsops,\ unsigned,\ nsops)
\]

The function semop() completes its task by calling do\_semtimedo\\p(). The function description of do\_semtimedop() is as follows:
\[
static\ long\ do\_semtimedop(int\ semid,
\]
\[
struct\ sembuf\ \_\_user\ *tsops, unsigned\ nsops,
\]
\[
const\ struct\ timespec64\ *timeout)
\]
To analyze the Linux failure mode caused by the wrong parameter of do\_semtimedop(), consider the following:
\begin{enumerate}
\item If the first parameter semid is incorrect, the following fault modes may occur:
“Semaphore does not exist error encountered while managing semaphore set”: When operating on a semaphore set, the semaphore does not exist. The function returns -EIDRM for “Identifier removed”.
\item If the second parameter tsops is incorrect, the following fault modes may occur:
“Illegal address error encountered when managing semaphore set”: When operating on a semaphore set, the address pointed to by the parameter sops or timeout is inaccessible. The function returns -EFAULT for “Bad address”.
\item If the third parameter nsops is incorrect, the following fault modes may occur:
“Number out of range error when managing semaphore set”: When operating on a semaphore set, the number in the specified semaphore set is out of range. The function returns -E2BIG for “Argument list too long”.
\end{enumerate}

\subsection{Linux Fault Mode Library}
We constructed a Linux fault mode library with 2870 fault modes by analyzing Linux source code and its 152 system calls. Each fault mode in the library has the following attributes: simulation\_method\\\_id, fault\_mode\_id, fault\_mode\_name, simulation\_method\_type and attach\_data. The simulation\_method\_id serves as a unique identifier for each fault mode, while the fault\_mode\_id denotes the type of each fault. The fault\_mode\_name provides a specific description of the fault mode, and the simulation\_method\_type indicates the type of simulation method used for this fault mode. Additionally, the attach\_data field contains various additional parameters that are required for injecting this fault mode.

\section{Fault Injection Tool based on the Linux Fault Mode Library}

FIFML is a fault injection tool that enables the analysis of Linux failures without modifying the OS source code, thereby eliminating the need to rebuild the code base. The structure of FIFML is illustrated in Figure 4, it comprises the control module, the fault mode library management module, the fault injection scheme generation module, the fault injection module, the log module, and the Linux fault mode library.

\begin{figure}
\centering
\includegraphics[width=0.5\textwidth]{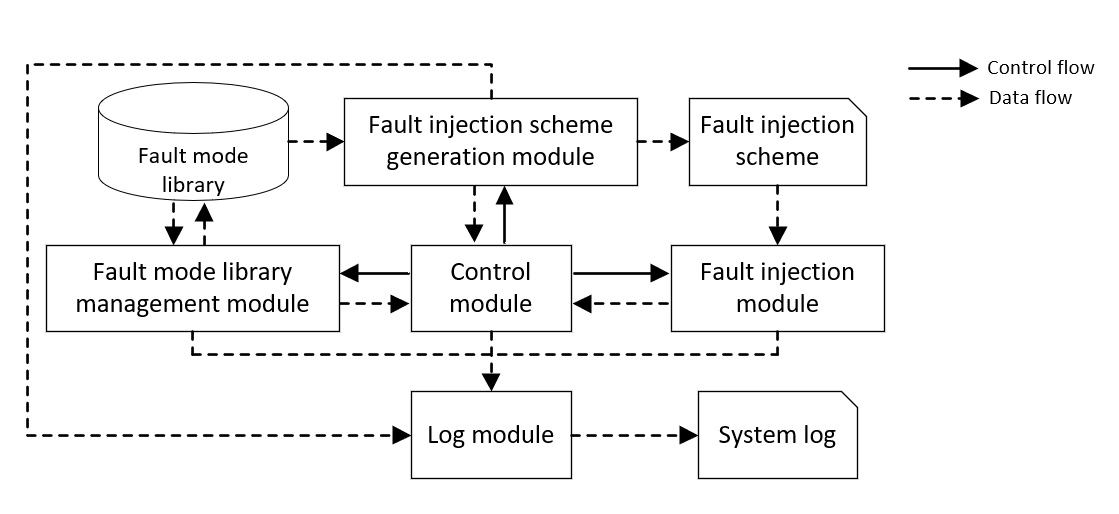}
\caption{\label{fig:fig4}FIFML Structure}
\end{figure}

\subsection{Fault Injection Scheme Generation Module}
The fault injection scheme generation module is responsible for receiving fault injection control information and generating a fault injection scheme based on acquired fault simulation method data. This module has two parts: the schema querier, which queries specific fault mode data from the fault mode library management module based on the control command, and the scheme generator, which generates the fault injection scheme using the fault mode data.

The fault injection scheme generation module executes the control command issued by the control module, generates a fault injection scheme based on the fault mode library and the control information, and sends it to the fault injection module. The fault injection scheme specifies the fault set to be simulated, the simulation method for each fault, the start time and duration of each fault, and the scope of influence of each fault. The generated fault injection scheme includes fault mode information, fault simulation method information, fault occurrence time, fault duration, target process PID, target file name, and other information. 

\subsection{Fault Injection Module}

The fault injection module implements the fault injection scheme and performs the specified fault simulation operation. This module is composed of three parts: the license checker, which verifies the validity of the tool license; the parameter checker, which ensures the legality of the received control commands; and the injector, which performs fault injection according to the fault injection scheme and manages the injected faults. Upon receiving a control command from the control module, the fault injection module interacts with the Linux kernel to execute the fault injection.

\subsection{Verification}
To validate the fault injection method, the fault mode library, and the injection tool, we conducted a series of experiments. We began by verifying the fault mode library's validity using tools to ensure that each item was effective and met the expected effect. In Section 3.A, the functional division of Step 1 ensures the theoretical completeness of the failure modes, and these fault modes were reviewed by our industry partners. Then, we conducted authenticity verification. We worked closely with industry partners to generate fault modes that completely covered a series of problems they actually encountered. Additionally, we analyzed a series of real failure scenarios submitted by the Anolis open-source community, they can be successfully simulated and reproduced.

We collected 786 bug reports submitted to the Anolis open-source community for the Anolis OS 8 series OS between December 2021 and April 2023. After screening the reports, we removed items that were not clearly described, caused by users' improper operations, or marked as errata. We then identified a set of 144 verifiable bugs, of which 74 could be simulated using our method. The remaining 70 bugs could not be simulated through fault injection, and were categorized as follows:

\textbf{(1) Missing packages and dependencies during installation:} There were 43 such errors, whose root cause lay in the installation source or other parts, and could not be simulated through fault injection. Although these errors are common in actual use, they can be quickly fixed by updating, so do not require failure analysis.
\begin{figure*}[!t]
\centering
\includegraphics[width=1\textwidth]{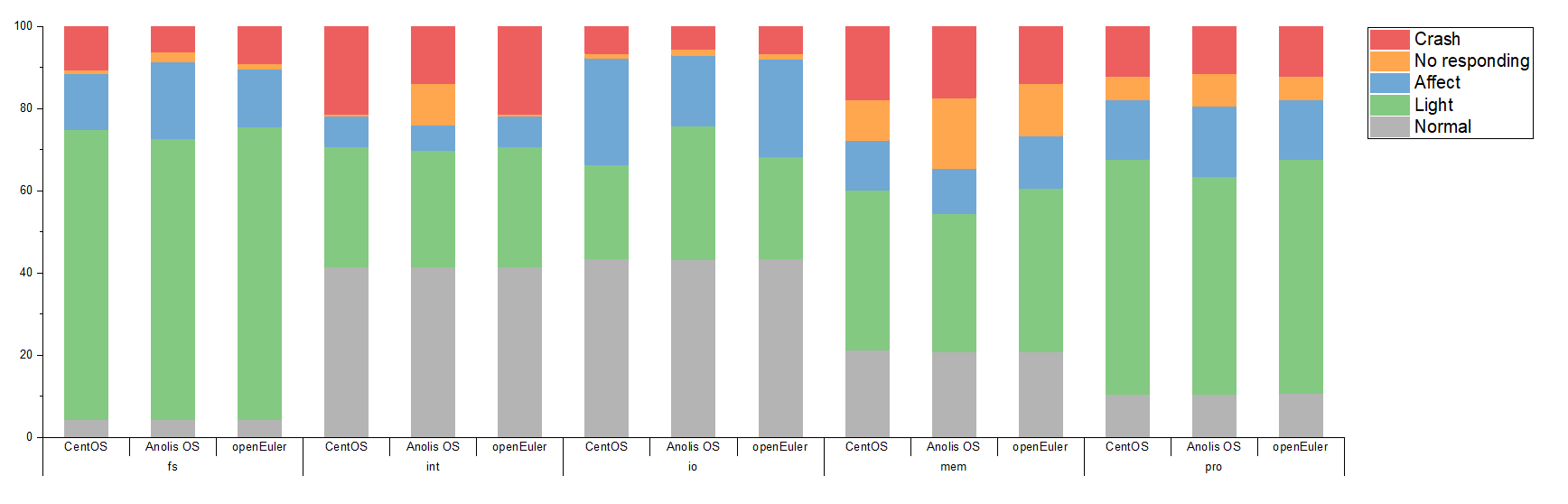}
\caption{\label{fig:fig5}Fault injection experimental results of three OSs}
\end{figure*}

\textbf{(2) User application failure:} There were 12 bugs in this category, such as garbled characters displayed in user applications or vulnerabilities in the Apache HTTP server, which could be resolved by using a newer version of the application. Although it is difficult to simulate faults in user applications using the fault injection based on internal functions targeting system calls, simulation targeting application program interfaces can be effective.

\textbf{(3) Hardware failure:} This type of error included 5 bugs, such as the inability to read the network card or an outdated BIOS version, whose root cause was hardware-related. Although fault mode simulations can help to observe and analyze the failure performance of upper layers, it is important to note that the fundamental problem lies in the hardware and should not be ignored.

\textbf{(4) Compatibility errors:} There were 10 bugs caused by adaptation or kernel compatibility, which could not be well simulated through fault modes .

Here is an example of fault simulation: Bug 1953 describes a Linux issue where pipe\_resize\_ring() lacks a lock. The function of pipe\_resize\_ring() is to adjust the size of the pipe ring. During the adjustment process, the content of the old pointer is copied to the new pointer and the space pointed to by the old pointer is released. Without the lock, another function post\_one\_notification() could cause a use-after-free error and result in an oops when inserting into the buffer. This scenario clearly exhibits the characteristics of kernel function failure, and fault injection into pipe\_resize\_ring() and post\_one\_notification() can directly simulate the failure. Furthermore, the fixed pipe\_resize\_ring() utilizes spin\_lock\_irq() and spin\_unlock\_irq() to control the lock. Fault injection into these functions can produce additional failure scenarios.

\section{Experiments and Analysis}
We conducted two types of experiments on three Linux-like OSs, i.e. CentOS-stream-8 (kernel 4.18.0-383.el8.x86\_64), AnolisOS-8.4-GA (kernel 4.18.0-305.an8.x86\_64) and openEuler-20.03-LTS-SP3 (kernel 4.19.90-2112.8.0.0131.oe1.x86\_64). The first type of experiment was the failure analysis based on fault injection, while the second was the performance analysis of the three OSs, both with and without fault injection. For each type of experiment, we ran Phoronix Test Suite as workload to warm up the system for 30 seconds. Phoronix includes a wide range of benchmarks covering different types of workloads, such as CPU, memory, disk, and graphics. The benchmarks are designed to stress-test the system and provide a comprehensive evaluation of its performance. We selected Phoronix as workload for fault injection because it provides a challenging and realistic set of workloads that can reveal potential failures in the system. Then, we performed fault injection using FIFML. Finally, we rebooted the system to its initial state before each experiment. We collected log files at the end of each experiment to determine whether the fault was successfully injected and whether it affected the OS performance. This experimental design enabled us to evaluate the robustness of the three OSs against faults and compare their performance with and without fault injection.

\subsection{The Experiment of Failure Analysis based on Fault Injection}
\subsubsection{Experiment Description}
The results of the failure analysis experiment are classified into five levels: (1) Crash: The system crashes after a fault is injected; (2) No Response: The system does not respond after a fault is injected; (3) Affect: Some processes are affected and cannot run normally after a fault is injected; (4) Light: After a fault is injected, the system can run but is affected, and some processes cannot provide services correctly; (5) Normal: The system runs normally after a fault is injected. For each experiment, a single fault is injected each time, and kernel functions from each of the three OSs are injected three times. Figure 5 shows the aggregated results of fault injection for the three OSs. Each bar in the figure represents the corresponding fault distribution.

For each OS, we performed 1250, 229, 621, 173, and 597 fault injection experiments on the modules of file system (fs), interrupt management (int), IO management (io), memory management (mem), and process management (pro), respectively. The number of fault injection experiments depends on the number of fault modes in the fault mode library. The proportion of file system and IO management modules that are seriously affected by injected faults (Crash or No Response) is relatively low, while the proportion of interrupt management module and process management module that are seriously affected by injected faults is relatively high. Comparing results show that Anolis OS performs well in file system and IO management, while openEuler performs better in memory management and process management. The three OSs performed similarly in interrupt management, but Anolis OS crashes less frequently.

\subsubsection{Experimental Analysis and Comparison}
In our experiments, we observed one type of failure related to the read file operation. We simulated IO errors, invalid parameter errors, and operation blocking by returning error codes such as EIO, EINVAL, and EWOULDBLOCK from the system call sys\_read(). Figure 6 shows an example of a system call returning an error code that leads to a runtime exception. CentOS and openEuler will crash directly without providing any meaningful information to the user, leading to a bad user experience. Anolis OS, on the other hand, will prompt the user that the library file cannot be read and provide a message of “Input/output error.”

\begin{figure}
\centering
\includegraphics[width=0.45\textwidth]{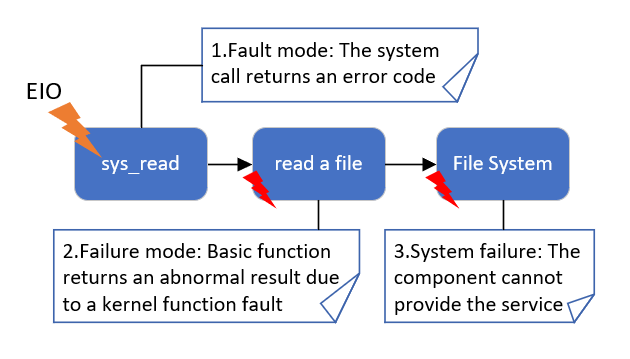}
\caption{\label{fig:fig6}A System Call Returns An Error Code Resulting in A Runtime Exception}
\end{figure}

Another type of function failure occurs during the operation of obtaining the file status according to the file descriptor. This failure is generated by returning EFAULT, ENOTDIR, and ELOOP from sys\_newfstat(). CentOS and openEuler cannot execute file operations and prompt the user that the library file cannot be loaded and gives Error 40 information. Some functions of Anolis OS, such as linking files produced during the compilation process, will prompt Error, but the overall performance of the system is stable. Anolis OS is more robust and can provide some help for subsequent failure recovery.

\subsubsection{Case Study}
In addition, some of the differences between various kernel versions of the components are highly related to the fault modes. Specifically, in the case of the sys\_munmap(), sys\_getdents64() and the remaining 8 system calls, a total of 39 fault modes have been identified that can cause certain kernel versions of the systemd components to fail, leading to core dumps and system crashes. Additionally, failures related to sys\_fadvise64() have similar behavior. To illustrate an example of failure analysis related to sys\_munmap(), we provided an analysis of the fault caused by the linux-mem-26-7-1 fault mode injected in our experiment. Upon injection, the system encounters an invalid parameter error while canceling the mapping of files or devices to memory, leading to a segmentation fault and system restart in Anolis. In contrast, the other two OSs behave normally. Further investigation using system log comparison reveals that the kernel was encountering issues while handling page faults triggered by sys\_munmap(). Specifically, in the experiment using the 4.18.0-305.an8 Anolis kernel, the page fault handling was found to be similar to that of the Linux 4.18.0-305.el8 kernel. Consequently, this similarity resulted in significant contention for the mmap\_sem lock. The situation became precarious when a fault was injected into the memory-mapping class, causing a core dump. Due to this injected fault, the process (often systemd) failed to release the mmap\_sem lock, leading to improper handling of the page fault. As a result, a “stack guard page was hit” error occurred, ultimately leading to a kernel panic due to stack overflow. This problem has been resolved in subsequent versions of the kernel.

The differences observed during the experiment and the corresponding analysis results were confirmed by our industry partners. We presented our findings in a report that was shared with the developers for their consideration. This feedback allowed the developers to gain valuable insights into the performance of their OS under various fault conditions, and helped to identify objects for improvement.

\subsection{Performance Analysis Experiment}
\subsubsection{The Selection of the Analysis Object}
To assess the impact of faults on Linux kernel performance, virtual file systems, interprocess communication systems, memory management systems, process management systems, and network interface modules of CentOS 8, Anolis OS, and openEuler are used as evaluation objects. Combined with the feature that FIFML can perform accurate fault injection for Linux kernel functions, the performance of their internal common functions before and after fault injection was measured, and the performance metric was represented by the delay in the completion of the function call.

In order to screen out the kernel functions and system calls that are frequently used in each object, we use strace \cite{b29} and ftrace \cite{b30} tools to build the call stack of Linux that runs continuously for a period of time under the working environment. The functions to be tested which are screened out according to the function calls in the stack involved 59 functions from five system modules.
\subsubsection{Performance Evaluation Metrics}
The performance analysis experiment obtains the time required for the execution of each function before fault injection and the time required for the execution of each function after fault injection ($TAF$), and the influence level of fault on the execution performance of the function was divided according to the difference in the time required for each operation before and after fault injection. In order to reasonably divide the influence level and reduce the impact of performance fluctuations, we defined a performance threshold ($PT$) for the execution of each function, which is used to estimate the maximum performance fluctuation of each function before fault injection. In addition, we define the performance standard deviation ($PSD$) of each function before fault injection. The worst performance ($WP$) of each operation before fault injection is obtained from the time required for each function to execute before fault injection. According to the 3 Sigma criterion in statistics, if the performance float follows the normal distribution, 99.73$\%$ of the data is within the range of 3 standard deviations of the mean. Therefore, in order to further improve the confidence level, under the condition of known standard deviations and the worst performance, the maximum allowed performance float is the sum of 3 times $PSD$ and $WP$, i.e.
\begin{equation}
PT=3\times PSD+WP\label{e1}
\end{equation}

According to the relationship between $TAF$ and $PT$, three levels of influence are defined

(1) No Influence:  $ TAF<PT$

(2) Mild Influence:  $PT \leq TAF < 5 \times PT$

(3) Serious Influence:  $ TAF\geq 5 \times PT$

“No Impact” indicates that there is little difference in operation performance before and after fault injection. “Mild Impact” indicates that operational performance deteriorates before and after fault injection, but the performance difference is considered tolerable by the user and has little impact on the performance of user processes and system services. “Serious Impact” indicates that operation performance deteriorates significantly before and after fault injection, and the performance of user processes and system services are affected.

We define the failure rate ($FR$), performance degradation rate ($PDR$), and performance level after fault injection ($PL_f$) to quantitatively describe the impact of fault injection on performance of CentOS 8, Anolis OS, and openEuler .

$PL_f$ depends on fault impact degree ($FID$). $FID$ is related to the number of faults that cause a serious impact ($NSF$), the number of faults that cause a mild impact ($NMF$), the number of faults that cause no responding ($NNR$), and the number of faults that cause system crash ($NOC$). But they have different impacts, In order to distinguish this difference, $NSF$,$NMF$,$NNR$ and $NOC$ are assigned different weights respectively.
\begin{equation}
FID=0.4\times NSF+1.0\times NMF+2.0\times NNR+3.0\times NOC\label{e2}
\end{equation}
The larger the $FID$, the more faults that affect performance, the higher the fault impact, the smaller the $PL_f$. And also, the larger the $FID$, the smaller the effect on $PL_f$. At the same time, $PL_f$ is also related to the total number of faults ($N$). So $PL_f$ can be expressed as
\begin{equation}
PL_f=e^\frac{-FID}{N} \times 100\%\label{e3}
\end{equation}
$FR$ means the probability that the system is in failure state and loses the ability to provide services. It depends on $N$, $NNR$ and $NOC$. The smaller the $FR$, the stronger the system's ability to maintain services after fault injection. So $FR$ can be expressed as
\begin{equation}
FR=(NNR+NOC)/N \times 100\%\label{e4}
\end{equation}
$PDR$ is the probability that a fault affects system performance and results in performance degradation (including loss of service capability). It depends on $N$, $NMF$, $NSF$, $NNR$, and $NOC$. A smaller $PDR$ indicates a smaller impact on system performance. So $PDR$ can be expressed as
\begin{equation}
PDR=(NMF+NSF+NNR+NOC)/N \times 100\%\label{e5}
\end{equation}
\subsubsection{Experimental Analysis and Comparison}
On the whole, Anolis OS performs better than CentOS and openEuler in virtual file systems, network interfaces, and process management systems. openEuler provides better inter-process communication performance than CentOS and Anolis OS, but provides more NSFS for virtual file systems and network interfaces than CentOS and Anolis OS.
 
Table 1 gives the number of faults that cause various impacts ($NMF$/$NSF$/$NNR$/$NOC$). The results show that, compared with CentOS and openEuler, Anolis OS has lower $NMF$, $NSF$ and $NOC$, but also has higher $NNR$, which indicates that more faults will cause Anolis OS to enter the non-responsive state and lose the ability to provide services. OpenEuler has the lowest $NOC$ and produces the fewest system crashes, but its larger $NMF$ and $NSF$ indicate that more failures may affect openEuler performance to some extent.
\begin{table}[tb]
\renewcommand{\arraystretch}{1.3}
\caption{Number of Faults That Cause Various Impacts ($NMF$ / $NSF$ / $NNR$ / $NOC$)}
\label{fault_injection_summary}
\centering
\begin{tabular}{cccccc}
\hline
System name & $NMF$ & $NSF$ & $NNR$ & $NOC$\\
\hline
CentOS 8 & 179 & 81 & 70 & 328\\

Anolis OS & 156 & 64 & 139 & 224\\

openEuler & 243 & 223 & 80 & 303\\

\hline
\end{tabular}
\end{table}

Based on the experimental data and the performance metrics defined above, we obtain $PL_f$, $FR$ and $PDR$ for CentOS 8, Anolis OS and openEuler. Table 2 gives $PL_f$, $FR$ and $PDR$ of the three OSs. Anolis OS obtains the best $PL_f$, $FR$ and $PDR$ scores because of lower $NMF$, $NSF$ and $NOC$. CentOS 8 lags behind Anolis OS in $PDR$ and $PL_f$. openEuler performs not so good on $PDR$ and $PL_f$  due to higher $NMF$, $NSF$ and $NOC$, but better on $FR$ than CentOS 8.
\begin{table}[tb]
\renewcommand{\arraystretch}{1.3}
\caption{Statistics of $FR$, $PDR$ and $PL_F$}
\label{fault_injection_summary}
\centering
\begin{tabular}{cccccc}
\hline
System name & $FR$ & $PDR$ & $PL_F$\\
\hline
CentOS 8 & 13.87$\%$ & 22.93$\%$ & 67.49$\%$\\

Anolis OS & 12.65$\%$ & 20.42$\%$ & 70.47$\%$\\

openEuler & 13.34$\%$ & 29.58$\%$ & 65.47$\%$\\

\hline
\end{tabular}
\end{table}
\subsubsection{Case Study}
Fault linux-mem-percpu-38-7-1 can simulate the fault of Linux system by causing \_\_alloc\_percpu() error while allocating dynamic percpu area. In the performance analysis experiment, this fault greatly increases the execution delay of kmem\_cache\\\_create() for CentOS and AnolisOS, but does not affect the execution delay of kmem\_cache\_create() for openEuler.

Therefore, we used ftrace to construct the call trees of three OSs, and analyzed the call chains involving kmem\_cache\_create() by contrast. CentOS and Anolis OS exhibit the same control flow. After executing find\_mergeable(), \_\_kmem\_cache\_alias() internally returns and continues to execute subsequent segments including \_\_alloc\_percpu(). Fault is triggered when \_\_alloc\_percpu() is executed, leading to system performance degradation. However, openEuler has a different control flow. \_\_kmem\_cache\_alias() internally executes sysfs\_slab\_alias after executing find\_mergeable(). And \_\_alloc\_percpu() is not included in subsequent program segments. Therefore, system performance is not affected.

After further analysis, it is found that the main reason for the above control flow difference is the inconsistency in the implementation of find\_mergeable() in the three OSs. In find\_mergeable(), CentOS and AnolisOS return NULL while openEuler returns a kmem\_cache structure on execution of the list\_for\_each\_entry\_reve\\rse segment. Inconsistencies in the returned results from find\_merge\\able() directly impact system performance discrepancies.

\section{Conclusion}
This paper presents a novel methodology for failure analysis of Linux-like OSs based on fault injection. We systematically define Linux-like fault modes by adopting the method of fault mode generation based on functional module division of Linux-like OSs. We construct a Linux fault mode library and develop a fault injection tool based on the fault mode library (FIFML). Then, we conduct fault injection experiments on three commercial Linux distributions, i.e. CentOS, Anolis OS and openEuler. To reasonably divide the influence level and reduce the impact of performance fluctuations, we introduce several performance metrics to measure the performance disparity of three OSs. The experimental results show that Anolis OS outperforms CentOS and openEuler in virtual file systems, network interfaces, and process management systems. These findings underscore the significance of our methodology in assessing OS reliability. By comprehensively examining various fault modes and their effects on performance, our methodology contributes to a better understanding of OS failure behavior and provides insights for future system optimization.

However, the completeness of the fault mode library is crucial for the effectiveness of FIFML in Linux failure analysis. Due to the high complexity of system failure behavior, the failure analysis time is long, and it is challenging to ensure the coverage of the fault mode library. 

As for future work, our first priority is to further improve the completeness of the Linux fault mode library. To reduce the manual effort required to construct a Linux fault mode library, we will combine practical experience with automated tools. Furthermore, we will employ fault injection based reliability testing to estimate OS reliability. We will explore the use of machine learning techniques to further enhance the accuracy and efficiency of our fault injection method. In addition, we will also extend the fault injection method to other levels (mainly user mode).

\end{document}